\documentclass{aipproc}
\def\selectedlayoutstyle{8x11single}
\layoutstyle\selectedlayoutstyle
\SetInternalRegister\hbadness{8000} 

\def\apj{\emph {ApJ}}
\def\apjs{\emph {ApJ Supp.}}
\def\apjl{\emph {ApJ Lett.}}
\def\mnras{\emph {MNRAS}}
\def\aap{\emph {A\&A}}
\def\aaps{\emph {A\&A Supp.}}
\def\araa{\emph {ARAA}}
\def\prl{\emph {PRL}}

\begin{document}
\title{Polarized Microwave Emission from Dust}
 \author{A. Lazarian}{
address={Department of Astronomy, University of Wisconsin, Madison, WI 53706},}
 \author{S. Prunet}{
address={Institut d'Astrophysique de Paris, 98bis Bld Arago, 75014 PARIS},
}
\copyrightyear{2001}

\begin{abstract}
Polarized emission from dust is an important foreground that
can hinder the progress in polarized CMB studies unless carefully
accounted for. We discuss potential difficulties associated with
the dust foreground, namely, the existence of different grain populations
with very different emission/polarization properties and variations of the
polarization yield with grain temperature. In this context we appeal for
systematic studies of polarized dust emission as the means of dealing
with this foreground.
\end{abstract}

\date{\today}

\maketitle

\section{Introduction}

Diffuse Galactic microwave emission carries important information on
the fundamental
properties of interstellar medium, but it also interferes with the
Cosmic Microwave Background (CMB) experiments
(see Bouchet et al. 1999, Tegmark et al. 2000). 
Polarization of the CMB provides information about the Universe
that is not contained in the temperature data. In particular, it offers a 
unique
way to trace specifically the primordial perturbations of tensorial nature
({\em i.e.} cosmological gravitational waves, see Seljak \& Zaldarriaga 1997, 
Kamionkowski et al. 1997),
and allows to break some important degeneracies that remain in the 
measurement of cosmological
parameters with intensity alone (Zaldarriaga et al. 1997, Davis 
\& Wilkinson 1999, 
Lesgourgues et al. 1999, Prunet et al. 2000).
Therefore, a number of groups
around the world (see Table~1 in Staggs et al. 1999) work hard to measure
the CMB polarization. In view of this work, the issue of determining the
degree of Galactic foreground polarization becomes vital.

Among different sources of polarized foregrounds, interstellar dust is probably
the most difficult to deal with. We can identify several reason for
that. First of all, dust has both a population of tiny grains (Leger \&
Puget 1984), which are frequently called PAH, along with the ``classical''
power-law distribution of larger grains (Mathis, Rumpl \& Nordsieck 1977).
Then the composition of grains changes with their size, which
influences both grain temperature and degree of grain alignment. Moreover,
both recent experience with microwave emissivity and theoretical studies
of expected polarization response (Draine \& Lazarian 1999)
show that the naive extrapolation of the grain properties from FIR to
microwave does not work. If we take into account that the very nature of
dust alignment that causes the polarization still remains somewhat
mysterious after more than half a century after its observational
discovery (see review by Lazarian 2000), the scope of the problem
becomes apparent.

The discovery of the anomalous emission in the range of 10-100~GHz
illustrates well the treacherous nature of dust.
Until very recently it has been thought that
there are three major components of the diffuse Galactic foreground:
synchrotron emission, free-free radiation from plasma (thermal bremsstrahlung)
and thermal emission from dust. In the microwave range of 10-90~GHz the latter
is subdominant, leaving essentially two
components. However, it is exactly in this range that an anomalous
emission was reported (Kogut et al. 1996a, 1996b). In the
paper by de Oliveira-Costa et al. (2000) this emission was nicknamed
``Foreground X'', which properly reflects its mysterious nature.
This component is spatially correlated with 100 $\mu$m thermal
dust emission, but its intensity is much higher than one can expect
by directly extrapolating thermal dust emission spectrum
to the microwave range. It is very likely that discoveries of
such a nature are expected when the foreground polarimetry is performed.

In this review, we briefly summarize what is known about the grain populations,
grain emission and grain alignment. We discuss the origin of the Foreground
X and its expected polarization.
Earlier reviews of the subject include (Prunet \& Lazarian 1999,
Draine \& Lazarian 1999 and Lazarian 2000).

\section{Observational Evidence}

\subsection{Infrared emission: extrapolation to microwave range}

Emission spectrum of diffuse interstellar dust was mostly obtained by
{\it InfraRed Astronomy Satellite} (IRAS) and infrared spectrometers
on the {\it COsmic Background Explorer} (COBE) and on the {\it InfraRed
Telescope in Space} (IRTS).

The emission at short wavelength, e.g. $<50$~$\mu$m,
arises from transiently heated very small grains. These grains have
so small heat capacity that the absorption of a single 6 eV starlight
photon rises their temperature to $T>200K$. Typically these grains
have less than $300$ atoms and can be viewed as large molecules
rather than dust particles. They are, however, sufficiently numerous
to account for $\sim 35\%$ of the total starlight absorption.
The contribution of those grains at the microwave frequencies
was thought to be negligible. 

In terms of CMB studies the most important is the emission from
cool classical dust.
Far infrared emission in the range from 1~mm (300~GHz) to
100~$\mu$m (3000~GHz) is primarily
due to dust particles heated by starlight to temperatures around
20~K. Those particles are ``classical'' grains known from ground-based
starlight absorption studies. It is convenient to approximate the 
far infrared emission with $\nu j_{\nu}$ peaking at $\lambda_m\sim
130$~$\mu$m and following the power-law corresponding the absorption
cross section $\sim \nu^{\beta}$, where $\beta\sim 1.7$,
 and temperature $T_{dust}=\frac{hc}{\lambda_m
k (4+\beta)}$ (see discussion in 
Draine \& Lazarian 1999). It was considered natural
to extrapolate this fit to frequencies
lower than 300~GHz, and no other contribution was expected from large
dust particles. 

If the extrapolation from infrared to microwave were as simple as it
is suggested above, dealing with dust
contribution would be trivial. Further research, however, 
revealed a much more complex
picture. Both classical and small grains were shown to be more
important microwave emitters than researchers used to assume. 
For tiny grains a new mechanism of emission was found (Draine \& Lazarian,
1998,a, henceforth DL98a, Draine \& Lazarian 1998b, henceforth DL98b),
while magnetic properties were shown to be important for microwave
emissivity of large grains (Draine \& Lazarian 1999).  
This example should be used to caution against simple minded
attempts to extrapolate polarization from infrared to microwave range.

\subsection{Anomalous microwave emission: unexpected discovery}

Until very recently it has been thought that
there are three major components of the diffuse Galactic foreground:
synchrotron emission, free-free radiation from plasma (thermal bremsstrahlung)
and thermal emission
from dust. In the microwave range of 10-90~GHz the latter
is definitely subdominant, leaving essentially two
components. However, it is exactly in this range that an anomalous
emission was reported (Kogut et al. 1996a, 1996b). In the
paper by de Oliveira-Costa et al. (2000) this emission was nicknamed
``Foreground X'', which properly reflects its mysterious nature.
This component is spatially correlated with 100 $\mu$m thermal
dust emission, but its intensity is much higher than one can expect
by directly extrapolating thermal dust emission spectrum
to the microwave range.

Since its discovery the Foreground X has been detected in the data
sets from Saskatoon (de Oliveira-Costa et al. 1997), OVRO (Leitch et
al. 1997), the 19~GHz survey (de Oliveira-Costa et al. 1998), and
Tenerife (de Oliveira-Costa et al. 1999, Mukherjee et al. 2000).
Initially, the anomalous emission was identified as thermal  bremsstrahlung
from ionized gas correlated with dust (Kogut et al. 1996a) and presumably
produced
by photoionized cloud rims (McCullough et al. 1999). This idea was
subjected to scrutiny in Draine \& Lazarian (1997) and criticized on
energetic grounds. Additional arguments against the free-free hypothesis
became available through correlating anomalous emission with ROSAT X-ray
C Band (Finkbeiner \& Schlegel 1999) and H$\alpha$ with 100~$\mu$m
emission (McCullough et al. 1999). They are summarized in Draine
\& Lazarian (1999). Recently
de Oliveira-Costa et al. (2000) used Wisconsin H-Alpha Mapper (WHAM)
survey data and established that the free-free emission ``is about an order
of magnitude below Foreground X over the entire range of frequencies
and latitudes where is detected''. The authors conclude that the
Foreground X cannot be explained as free-free emission. Additional
evidence supporting this conclusion have come from
a study at 5, 8 and 10~GHz by Finkbeiner, Schlegel , Frank \& Heiles (2001).

The spectrum of the Foreground X is not consistent with
synchrotron emission, and maps at 408 MHz (Haslam 1981) and
1.42 GHz (Reich \& Reich 1988)
do not correlate with the observed 15-100 GHz intensity,
so the anomalous emission is evidently not synchrotron radiation from
relativistic electrons.

Correlations of the Foreground X with dust induced Draine \& Lazarian (1998a,b,
1999) to conjecture that it can be indeed due to dust. It is encouraging
that the observational evidence obtained
since the theoretical predictions were published has supported
the theory.  

\subsection{Polarization from dust: half a century puzzle}

Polarization due to interstellar dust alignment was discovered in the middle of
the last century (Hiltner 1949, Hall 1949) and was studied initially via
starlight extinction and more recently through emission. Correlation of
the polarization with the interstellar magnetic field revealed that electric
vector of light polarized via starlight extinction tend to be parallel
to magnetic field\footnote{The polarizations in emission and in extinction
are orthogonal if they are produced by the same grains.}. This corresponds to grains being aligned with their
longer axes perpendicular to the local magnetic field. Due to the
presence of the stochastic magnetic field, the polarization patterns
are pretty involved.

The existing
data presents a complex picture. It is generally accepted that the observations
indicate that the 
ability to produce polarized light depends on grain size and
grain composition. For instance, a limited UV polarimetry dataset available
indicates that graphite grains tend not to be aligned 
(see Clayton et al. 1997),
while maximum entropy technique applied to the existing data
by Martin \& Kim (1995) show that large $>6\times 10^{-6}$~cm
grains are responsible for the polarization via extinction.

Moreover, the environment of grains seems to matter a lot (Goodman 1995,
Lazarian, Goodman \& Myers 1997). A study by Arce et al. (1998) indicates
that grains selectively extinct starlight up to optical depth $A_v<3$.
Recent emission studies (Hildebrand et al. 1999, 2001) produced a polarization spectrum
for dense clouds that reveal a tight correlation between grain temperature
and its ability to emit polarized light. As multicomponent fits
invoking grains of different temperature
were claimed to provide a better fit for the observed 1~mm-100~$\mu$m
emission (see Finkbeiner, Schlegel \& Davis 1999), this correlation may be 
very troublesome
for the attempts to construct polarization templates.

\section{Polarized Emission from Classical Dust}

\subsection{Grain alignment: light at the end of the tunnel?}

The basic explanation of polarized radiation from dust is straightforward.
Aligned dust particles preferentially extinct (i.e. absorb and scatter)
the $E$-component of starlight parallel to their longer axis.
Thermal $E$-component of the emitted radiation, on the contrary,
is higher along the
longer axis. Thus for aligned grains one must have polarization.
What is the cause of alignment?

Grain alignment is an exciting and very rich area of reseach. For example,
two new solid state effects have been discovered recently in the process
of understanding grain dynamics (Lazarian \& Draine 1999, 2000).
It is known that a number of 
mechanisms can provide grain alignment (see review by Lazarian
2000 and Table~1 in Lazarian, Goodman, \& Myers 1997). Some of
them rely on paramagnetic dissipation of rotational energy 
(Davis-Greenstein 1951, Purcell 1979, Mathis 1986, Lazarian \& Draine 1997,
Lazarian 1997a, Roberge \& Lazarian 1999) , some
appeal to the anisotropic gaseous bombardment when a grain moves
supersonically through the ambient gas (Gold 1951, Purcell \& Spitzer 
1971, Dolginov \& Mytrophanov 1976, Lazarian 1994, 1997b, Roberge,
Hanany \& Messinger 1995, Lazarian
\& Efroimsky 1996). Grains are definitely paramagnetic and sometimes
even strongly magnetic. Supersonic grain motions may be due to
outflows (Purcell 1969),
Alfvenic turbulence (Lazarian 1994) or ambipolar diffusion (Roberge
\& Hanany 1990).

At present,
grain alignment
via radiative torques (Draine \& Weingartner 1996, 1997)
looks preferable, although the theory and the understanding of the mechanism
are far from being complete. The mechanism appeals to a spin-up of a grain
as it differentially scatters left and right polarized photons (
Dolginov 1972, Dolginov \& Mytrophanov 1976). 
This process acts efficiently if the irregular
grain has its size comparable with the photon wavelength. The mechanism
can account for the systematic variations
of the alignment efficiency with extinction. 
 
However, other mechanisms should
also work. For instance,
paramagnetic mechanism may preferentially act on small grains (Lazarian
\& Martin 2002), while
mechanical alignment may act in the regions of outflows (Rao
et al. 1998). In general, the variety of Astrophysical conditions allows
various mechanisms to have their niche. 

Note, that in interstellar circumstances grain alignment
happens in respect to magnetic field, even if the mechanism of alignment
is not of magnetic nature. This is due to the fact that
the Larmor precession of grains is so fast compared to the time scales
over which either magnetic field changes its direction or the alignment
mechanism acts. In general, the alignment 
may happen both parallel and perpendicular to magnetic field. 
In most cases, the alignment happens with long grain axes perpendicular
to magnetic field, however. 

The history of grain alignment research is full of surprises. Initially
it looked so ubiquitous that observers were not even interested in the theory
of alignment. But then it showed that it may fail within molecular clouds.  
What will be the next surprise?

\subsection{Diffuse gas and molecular clouds: different beasts?}

Alignment of grains is different in diffuse gas and molecular clouds.
Lazarian, Myers \& Goodman (1997) showed that in dark clouds
without star formation all alignment mechanisms fail. Indeed,
grain alignment depends on non-equilibrium processes\footnote{To avoid
confusion we should remind the reader that interstellar grain
alignment is very different from the alignment of ferromagnetic particles
in the external magnetic field. The latter is the equilibrium process with
the align particles corresponding to the lowest energy level of the system.
The grain alignment is a {\it dissipative} process that requires constant
driving and vanishes in thermodynamic equilibrium.}, while interiors
of dark clouds are close to thermodynamic equilibrium.

As soon as stars are born within clouds, the conditions in their vicinity
become favorable for grain alignment. This explains why far infrared
polarimetry detects aligned grains, while near infrared and optical polarimetry
fails. The latter point is a subject of controversy. In the recent paper
by Padoan et al (2001) it is claimed that far infrared polarimetry does
not provide us with any new information compared with optical and near-infrared
studies. We are worried about this conclusion. Indeed, if anything, radiative
torques must be active near the newly born stars and the spectropolarimetric
studies of Hildebrand et al. (2001) indicate the existence of aligned hot
grains. These aligned grains are selectively warmer and
should reveal the structure of magnetic field in the star cruddle
after the star is born. This information is unlikely to be obtained via
short-wave polarimetry. 
Results by Padoan et al. (2001) may be
relevant, however, to 850~$\mu$m
polarization observed by Ward-Thompson et al. (2000) from dense pre-stellar
cores where radiative torques must be inefficient\footnote{Alternatives
are discussed in Lazarian (2000).}.    

We may hope that grain alignment in diffuse clouds is more uniform.
Radiation freely penetrates them and therefore the radiative torques
must ensure good alignment.
This assumption was used in Fosalba et al (2002)
who related the polarization from dust extinction and the polarization
from dust emission. Further research in this direction is necessary.

\subsection{Complications: turbulence, heating ...}

Interstellar medium is very complex and this tells on polarization.
As we have discussed earlier, grain alignment traces the direction of
the local magnetic field.
In the presence of turbulence, this field is very complex. The resulting
polarization depends on the telescope resolution at a particular
wavelength. A possible way of dealing with this complication is to
correct for the field stochasticity. Tensor description 
of turbulent magnetic field was 
obtained in Cho, Lazarian \& Vishniac (2002) and this can be used for the
purpose.  

Earlier on we mentioned that radiative torques may be responsible for the
bulk of grain alignment. As starlight also heats the grains
the systematic variations in the
alignment efficiency are expected 
for grains of different temperatures. Moreover, radiative torques depend on
grain size and grain composition and so do grain temperatures.
These and related issues
require a further study and a further work on the radiative
torque theory is necessary. 

It is unfortunate for the CMB research
 that we still do not understand many processes related
to the polarization arising from classical grains. The good news, however, is
that we will have to understand those processes along with the structure
of Galactic magnetic field at high latitude if we ever want to understand the
CMB polarization well. As the bonus from this research we will get an
insight into the operation of Galactic dynamo, high latitude MHD turbulence,
turbulent mixing and will make many yet unforeseen discoveries.

\section{Polarized Emission from Spinning Dust}

Can the ultrasmall grains observed via Mid-IR be important at the microwave
range? The naive answer to this question is no, as the total mass in those
grains is small.

However, DL98a appealed to a different mechanism of emission,
namely, to the rotational emission\footnote{The very idea of grain
rotational emission was first
discussed by Erickson (1957). More recently,
after the discovery of the population of ultrasmall grains,
Ferrara \& Dettmar (1994) noted that the rotational emission from such
grains may be observable, but their treatment assumed Brownian
thermal rotation of grains, which is incorrect.}
that must emerge when a grain with a dipole moment $\mu$  rotates
with angular velocity $\omega$.

For the model with the most
likely set of parameters, DL98a obtained a reasonable fit with observations
available at that time. It is extremely important that new data points
obtained later (de Oliveira-Costa et al. 1998,
de Oliveira-Costa et al. 1999) correspond to the already published
model. The observed flattening of the spectrum and its turnover
around $20$~GHz agree well with the spinning dust predictions.

Microwave emission from spinning grains is expected to be polarized if
grains are aligned. Alignment of ultrasmall grains which are
essentially large molecules is likely to be different from alignment
of large (i.e. $a>10^{-6}$~cm) grains.
One of the mechanisms that might produce alignment of the ultrasmall
grains is the
paramagnetic dissipation mechanism
by Davis and Greenstein (1951). The Davis-Greenstein alignment mechanism is
straightforward: for a spinning grain
the component of interstellar magnetic field
perpendicular to the grain angular velocity varies in grain coordinates,
resulting in time-dependent magnetization, associated
energy dissipation, and a torque acting on the grain.
As a result grains  tend to rotate
with angular momenta parallel to the
interstellar magnetic field.

Lazarian \& Draine (2000, henceforth LD00) found
that the traditional picture
of paramagnetic relaxation is
incomplete, since it
disregards the so-called ``Barnett magnetization'' (Landau \& Lifshitz 1960).
The Barnett effect, the inverse of the Einstein-De Haas effect,
consists of the spontaneous magnetization of
a paramagnetic body
rotating in field-free space. This effect can be understood in
terms of the lattice sharing part of its angular momentum with
the spin system. Therefore the implicit assumption in Davis
\& Greenstein (1951)--
that the magnetization within a {\it rotating grain} in a {\it static}
magnetic field is equivalent to the magnetization within a
{\it stationary grain} in a {\it rotating} magnetic field --
is clearly not exact.

LD00 accounted for the ``Barnett magnetization'' and termed the effect
of enhanced relaxation arising from grain magnetization ``resonance
relaxation''. It is clear from Fig.~1 that resonance relaxation persists
at the frequencies when the Davis-Greenstein relaxation vanishes. However
the polarization is marginal for $\nu>35$~GHz anyhow. The discontinuity
at $\sim 20$~GHz
is due to the assumption that smaller grains are planar, and larger
grains are spherical. The microwave emission will be polarized
in the plane perpendicular to magnetic field.

\begin{figure}
\resizebox{\textwidth}{12cm}{\includegraphics{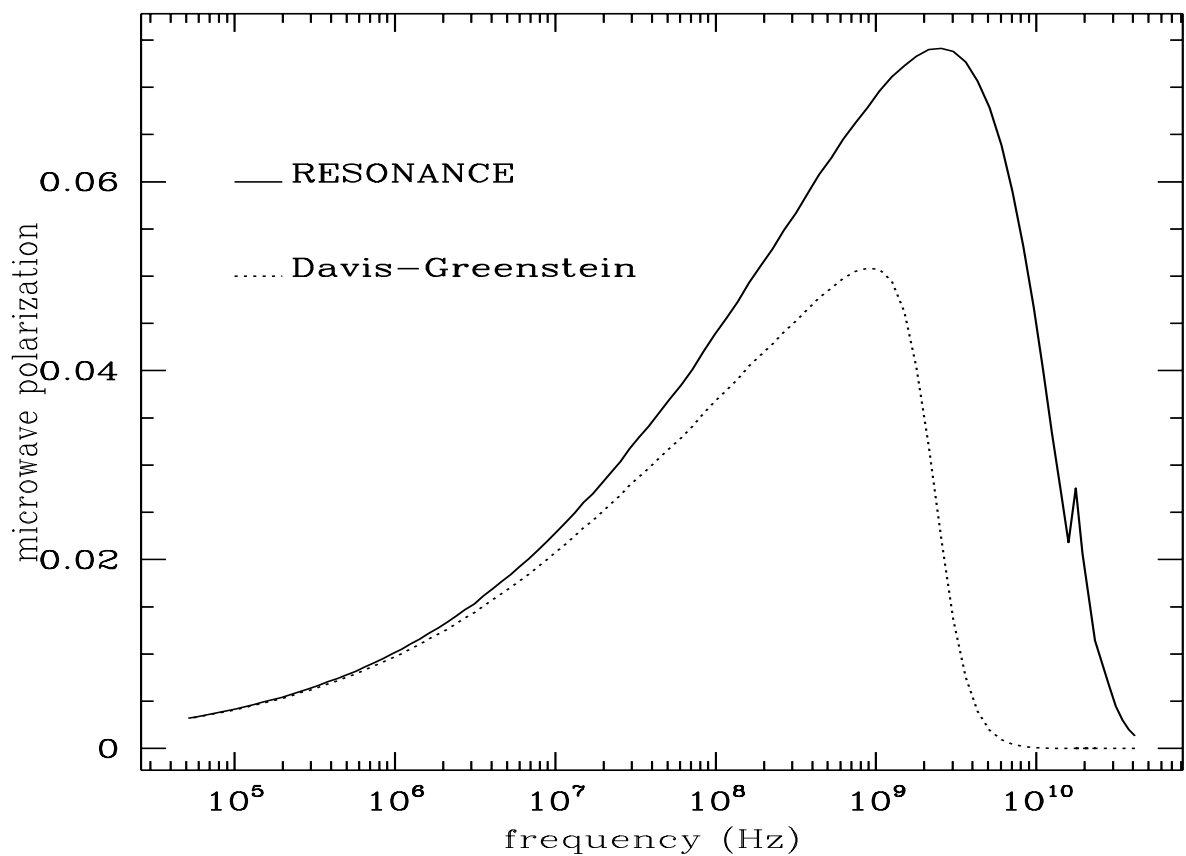}}
\caption{Polarization for both
        resonance relaxation and Davis-Greenstein relaxation for grains in
        the cold interstellar medium as a function of frequency (from LD00).
For resonance relaxation the saturation effects
(see eq.\ (1)) are neglected, which means
that the upper curves correspond to the {\it maximal} values allowed by the
paramagnetic mechanism.}
\end{figure}

Can we check the alignment of ultrasmall grains via infrared polarimetry?
The answer to this
question is ``probably not''. Indeed, as discussed earlier,
infrared emission from ultrasmall grains, e.g. 12 $\mu$m emission,
takes place as grains absorb UV photons. These photons raise
grain temperature, randomizing grain axes in relation to
its angular momentum (see Lazarian \& Roberge 1997). Taking values
for Barnett relaxation from Lazarian \& Draine (1999), we get
the randomization time of the $10^{-7}$~cm grain to be
 $2\times 10^{-6}$~s, which is less than grain cooling time. As the
result, the emanating infrared radiation will be polarized very marginally.
If, however, Barnett relaxation is suppressed, the randomization time
is determined
by inelastic relaxation (Lazarian \& Efroimsky 1999) and is
$\sim 0.1$~s, which would entail a partial polarization of
infrared emission.

\section{Polarized Emission from Magnetic Grains}

While the spinning grain hypothesis got recognition in the
community, the magnetic dipole emission model suggested by Draine
\& Lazarian (1999, henceforth DL99) was left essentially unnoticed.
This is unfortunate, as magnetic dipole emission provides a
possible alternative explanation to the Foreground X. Magnetic
dipole emission is negligible at optical and infrared frequencies.
However, when the
frequency of the oscillating magnetic field approaches the precession
frequency of electron spin in the field of its neighbors, i.e.
$10$~GHz, the magneto dipole emissivity becomes substantial.

How likely is that grains are strongly magnetic? Iron is the fifth
most abundant element by mass and it is well known that it resides in
dust grains (see Savage \& Sembach 1996). If $30\%$ of grain mass is
carbonaceous, Fe and Ni contribute approximately $30\%$ of the remaining
grain mass. Magnetic inclusions are widely discussed in grain
alignment literature (Jones \& Spitzer 1967, Mathis 1986, Martin 1995,
Goodman \&
Whittet 1996).
If a substantial part of this material is ferromagnetic
or ferrimagnetic, the magneto-dipole emission can be comparable to that
of spinning grains. Indeed, calculations in DL99 showed that less than
$5\%$ of interstellar Fe in the form of metallic grains or inclusions
is necessary to account for the Foreground X at 90~GHz, while magnetite,
i.e. Fe$_3$O$_4$,
can account for a considerable part of the anomalous emissivity over
the whole range of frequencies from 10 to 90~GHz. Adjusting the
magnetic response of the material, i.e. making it more strongly magnetic
than magnetite, but less magnetic than pure metallic Fe, it is possible
to get a good fit for the Foreground X (DL99).

How can magneto-dipole emission be distinguished from that from
 spinning grains?
The most straightforward way is to study microwave emission from regions
of different density. The population of small grains is
depleted in dark clouds (Leger and Puget 1984)
and this should result in a decrease of
contribution from spinning grains.
Private communication from Dick Crutcher who attempted such measurements
corresponds to this tendency, but the very detection of microwave
emissivity is a 3$\sigma$ result.
Obviously the corresponding  measurements are
highly desirable.
As for now,
magnetic grains remain a strong candidate process for producing
part or even all of Foreground X.
In any case, even if magnetic
grains provide subdominant contribution, this can be important
for particular cases of CMB and interstellar studies. For instance,
polarization from magnetic grains may dominate that from spinning
grains even if the emission from spinning grains is more of higher
level.

The mechanisms of producing polarized
magneto-dipole emission is similar to that
producing polarization of electro-dipole  thermal emission
emitted from aligned non-spherical grains (see Hildebrand 1988).
There are two
significant differences, however. First, strongly magnetic
grains can contain just a single magnetic domain. Further magnetization
along the axis of this domain is not possible and therefore the
magnetic permeability of the grains gets anisotropic: $\mu=1$ along
the domain axis, and $\mu=\mu_{\bot}$ for a perpendicular direction.
Second, even if a grain contains tiny magnetic inclusions and can be
characterized by isotropic permeability, polarization that it produces
is orthogonal to the electro-dipole radiation emanating through
electro-dipole vibrational emission. In case
of the electo-dipole emission, the longer grain axis defines the vector
of the electric field, while it defines the vector of the
magnetic field in case
of magneto-dipole emission.

The results of calculations for single domain iron particle (longer axis
coincides with the domain axis) and a grain with metallic Fe inclusions
are shown in Fig.~2. Grains are approximated by ellipsoids $a_1<a_2<a_3$
with ${\bf a_1}$ perfectly aligned
parallel to the interstellar magnetic field ${\bf B}$. The polarization
is taken to be positive when the electric vector of emitted radiation
is perpendicular to ${\bf B}$; the latter is the case for electro-dipole
radiation of aligned grains. This is also true (see Fig.~3) for high
frequency radiation from single dipole grains. It is easy to see
why this happens. For high frequencies $|\mu_{\bot}-1|^2\ll 1$
and grain shape factors are unimportant. The only important thing is
that the magnetic fluctuations happen perpendicular to ${\bf a_1}$.
With ${\bf a_1}$ parallel to ${\bf B}$, the electric fluctuations
tend to be perpendicular to ${\bf B}$ which explains the polarization
of single domain grain being positive. For lower frequencies magnetic
fluctuations tend to happen parallel to the intermediate size axis
${\bf a_2}$. As the grain rotates about ${\bf a_1}\|{\bf B}$,
the intensity in a given direction reaches maximum when an observer
sees the ${\bf a_1} {\bf a_2}$ grain cross section. Applying earlier
arguments it is easy to see that magnetic fluctuations are parallel
to ${\bf a_2}$ and therefore for sufficiently large $a_2/a_1$ ratio
the polarization is negative. {\it
The variation of the polarization direction with
frequency presents the characteristic signature of magneto-dipole emission
from aligned single-dipole grains and it can be used to separate this
component from the CMB signal}. Note that the degree of polarization is
large, and such grains may substantially interfere with the attempts
of CMB polarimetry. Even if the
intensity of magneto-dipole emission is subdominant
to that from rotating grains, it can still be quite important in
terms of polarization.
A relatively weak polarization response is expected for grains with
magnetic inclusions (see Fig.~2). The resulting emission is negative
as magnetic fluctuations are stronger along longer grain axes, while
the short axis is aligned with ${\bf B}$.

\begin{figure}
\resizebox{\textwidth}{12cm}{\includegraphics{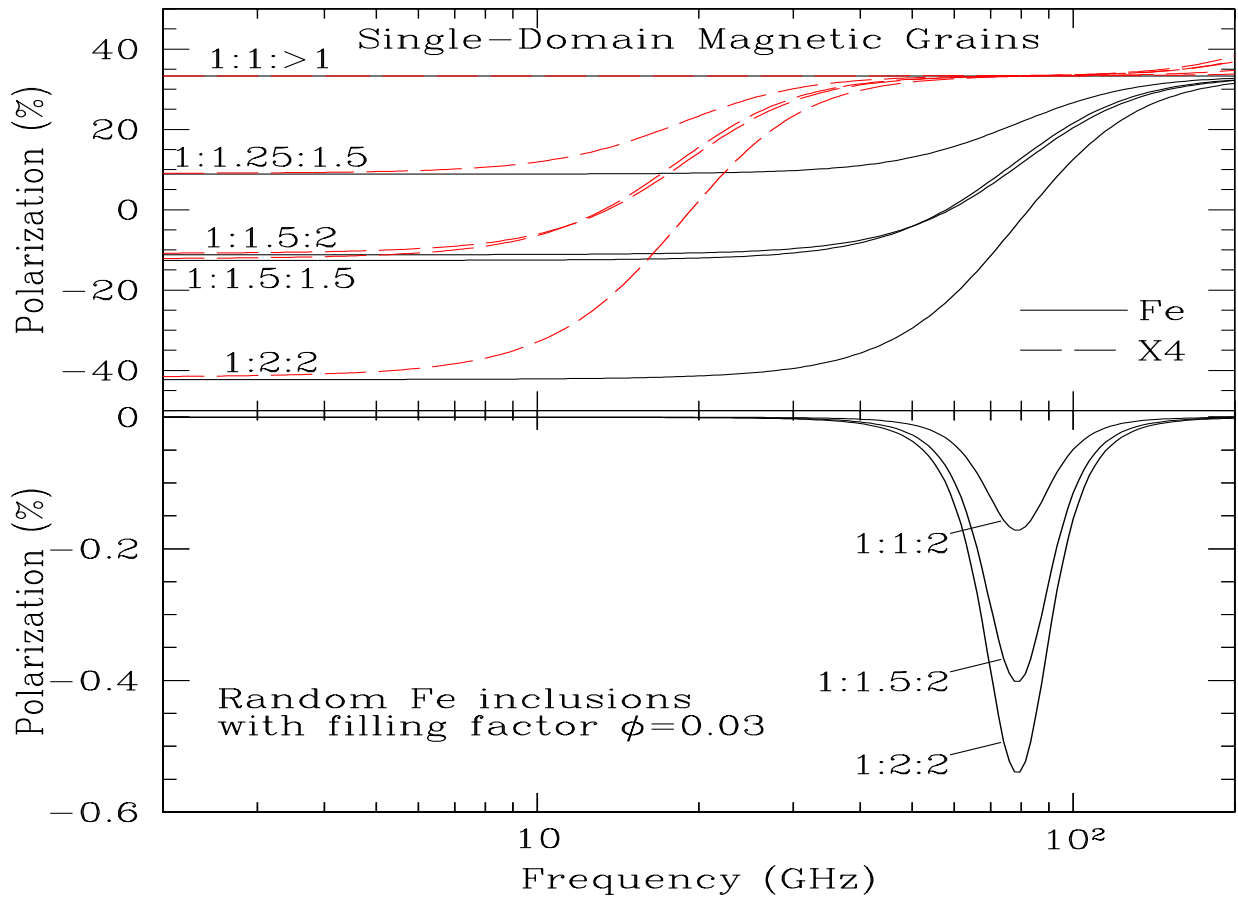}}
\caption{ Polarization from magnetic grains (from DL99). Upper panel:
Polarization of thermal emission from perfectly aligned single
domain grains of metallic Fe (solid lines) or hypothetical magnetic
material that can account for the Foreground X (broken lines).
Lower panel: Polarization from perfectly aligned grains with
Fe inclusions (filling factor is 0.03). Grains are ellipsoidal and
the result are shown for various axial ratios.
}
\end{figure}

Systematic studies of dust foreground polarization should improve
our insight into the formation dust, its structure, its composition.
For instance, DL99 showed that the present-day microwave measurements
do not allow more than 5\% of Fe to be in the form of metallic iron.
More laboratory measurements of microwave properties of candidate
materials are also necessary. Some materials, e.g. iron, were studied
at microwave range only in the 50's and this sort of data must be checked
again using modern equipment.

\section{Polarized dust emission as a Foreground to CMB measurements}

As we have seen in the previous sections, dust emission at microwave and submillimeter
frequencies is likely to be polarized, thus representing a foreground emission to 
the cosmic microwave background polarized (hereafter CMBP) emission.
In order to estimate the impact of this foreground on the CMBP measurements,
it is essential to compare not only their relative dominance in different
frequency channels, but also their spatial statistical distribution since 
the CMBP emission is expected to have a very distinctive spatial distribution.
It is therefore important to compare foreground and background emission at different
wavelengths and \emph{at different scales}. 
In the following we will concentrate on frequencies where the thermal emission of 
dust is dominant over the rotational (or magnetic dipole) emission. This corresponds in
particular to the frequency range covered by the High Frequency Instrument aboard
the Planck satellite, which offers the highest possible sensitivity to the tiny 
CMBP emission.

\subsection{Depolarization factors}

Given the low optical depth of the diffuse ISM (where the CMBP measurements will be made, 
i.e. at high Galactic latitude), the polarized emission of dust for a particular line of sight 
depends on the integrated signal throughout the entire emitting medium. 
Even if the grains are perfectly aligned with their short axis along the magnetic field
lines, the magnetic field lines reversals will act as a depolarization mechanism on the
observed dust emission. The statistical distribution of the polarized emission is therefore
the result of a complicated interplay between the dust density distribution, possibly its
temperature distribution (resulting in varying alignment efficiencies, as previously discussed), 
and the magnetic field distribution (direction and amplitude 
a priori).
Following Greenberg(1968), we will define a depolarization factor $\Phi$ which gives the
relation between the intrinsic dichroic polarised cross-section of a grain and the 
(average) observed polarized cross-section along a typical line of sight. This factor is
defined as follows:
\begin{eqnarray}
\Phi &=& RF\cos\gamma \\
R &=& {3\over 2}(\langle\cos^2\beta\rangle-{1\over 3}) \\
F &=& {3 \over 2} (\langle\cos^2\theta\rangle-{1\over 3})
\end{eqnarray}
where $\beta$ and $\theta$ are the angle between the grain major axis (axis of maximal inertia)
and the direction of the magnetic field, and the angle between the random (turbulent) component of the magnetic
field and the regular component respectively; finally $\gamma$ is the angle between the regular component
of the magnetic field with the plane of the sky.
In principle the Rayleigh reduction factor $R$ is decomposed into the partial alignment of the grain
angular momentum ${\bf J}$ with its major inertia axis and the partial alignment of ${\bf J}$ with the
magnetic field. However the statistical distributions of these two angles are generally not independent
(see for instance Roberge \& Lazarian 1999) and it is precisely the determination of their combined
statistical average $R$ that is the goal of grain alignment theory.
In this section, we will assume a ``worst case scenario'' (in terms of CMBP measurement contamination)
by assuming that the grain alignment mechanism is perfect (hence $R=1$).
We are then left with the depolarization due to magnetic field line warping, and the value of the polarized
cross-section which depends on grain composition, size and axis ratio (we assume grains of spheroidal shape,
see e.g. Lee \& Draine 1985). 

\subsection{Polarized cross-sections}

The polarized cross-sections in the infrared wavelengths are primarily a function of the
grain composition and shape, and are not very sensitive to the size distributions of the 
grains provided that the wavelengths considered are much larger than the maximal grain sizes
considered. In particular, the spectro-polarimetric properties of grains around absorption
features of their component materials can be used to constrain the grain axis ratios in the
assumption of spheroidal shapes (see e.g. Lee \& Draine 1985, Hildebrand \& Dragovan 1995).
For instance, prolate and oblate spheroidal grains will have different locations of the peak
polarization in absorption around the $9.7 \mu m$ silicate feature. The $3.1 \mu m$ ice feature
can be used also to distinguish between prolate and oblate grains for a fixed grain core composition.
Both Lee \& Draine (1985) and Hildebrand \& Dragovan (1995) find that oblate grains with axis ratios
in the range 1:2 and 2:3 can reproduce the observations reasonably well.
For the mixture of silicate and graphite grains needed to reproduce the $9.7 \mu m$ and $2.2 \mu m$ absorption
features, and for a given grain shape one can compute the polarization efficiency in the millimeter
range to be of order $\sim 35 \%$ for a perfectly aligned grain in the plane of the sky 
(Hildebrand \& Dragovan 1995).
The difference between this theoretical upper limit of polarization efficiency and the observed histograms
of polarization degree in emission at $100\mu m$ (peak at $2\%$ and maximum around $9\%$)
toward the Orion nebulae would then be explained by the depolarization factors described in the preceding section,
i.e. the magnetic field lines entanglement and the partial alignment of grains.
To estimate these factors in the diffuse ISM, we will follow a pragmatic approach described in Prunet et al. (1998).

\subsection{Estimating the polarization efficiency in emission: a first model}

\begin{figure}
\label{fig:histogram}
\resizebox{10cm}{!}{\includegraphics{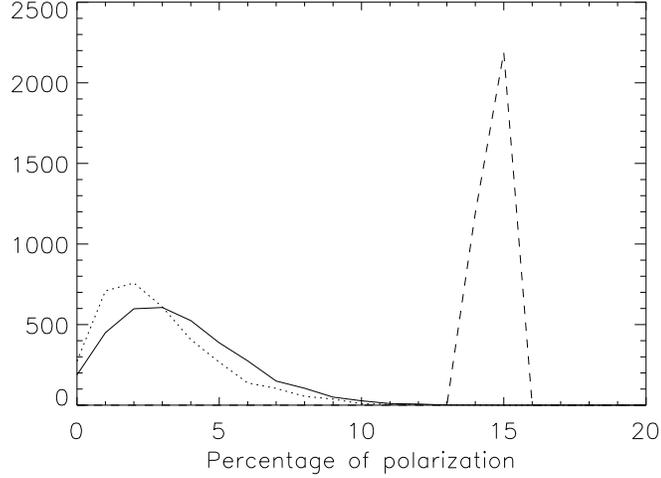}}
\caption{Predicted histograms of Galactic dust polarized degree in emission for the three different assumptions 
for magnetic field lines distribution (see text). The solid, dotted and dashed lines correspond respectively to the three
cases described in the text. While two of the assumptions can reproduce the observations
reasonably, the constant magnetic field case is clearly rejected.}
\end{figure}

\begin{figure}
\label{fig:spectra}
\begin{tabular}{c}
\mbox{\includegraphics[width=\textwidth/3]{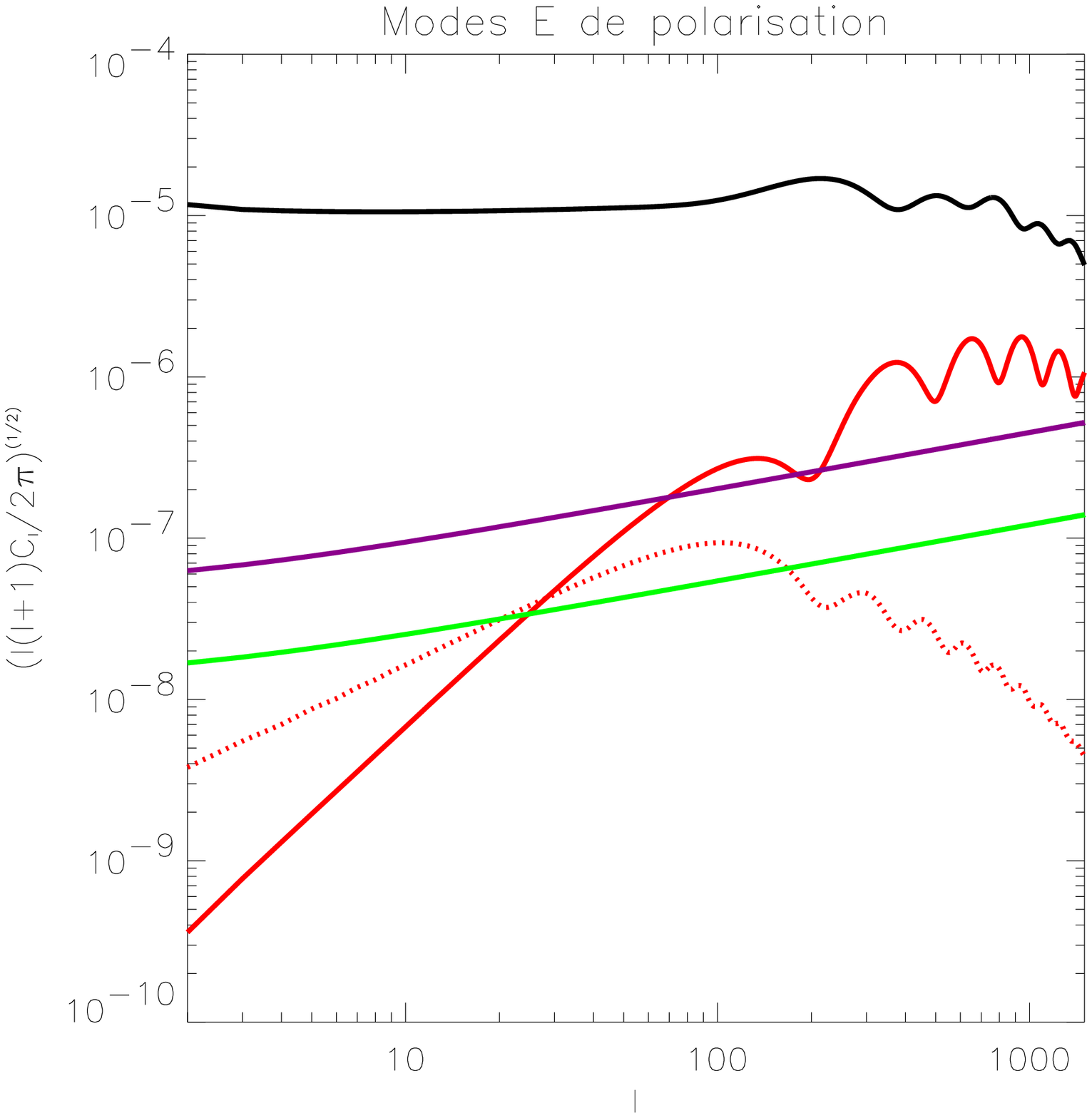}}
\mbox{\includegraphics[width=\textwidth/3]{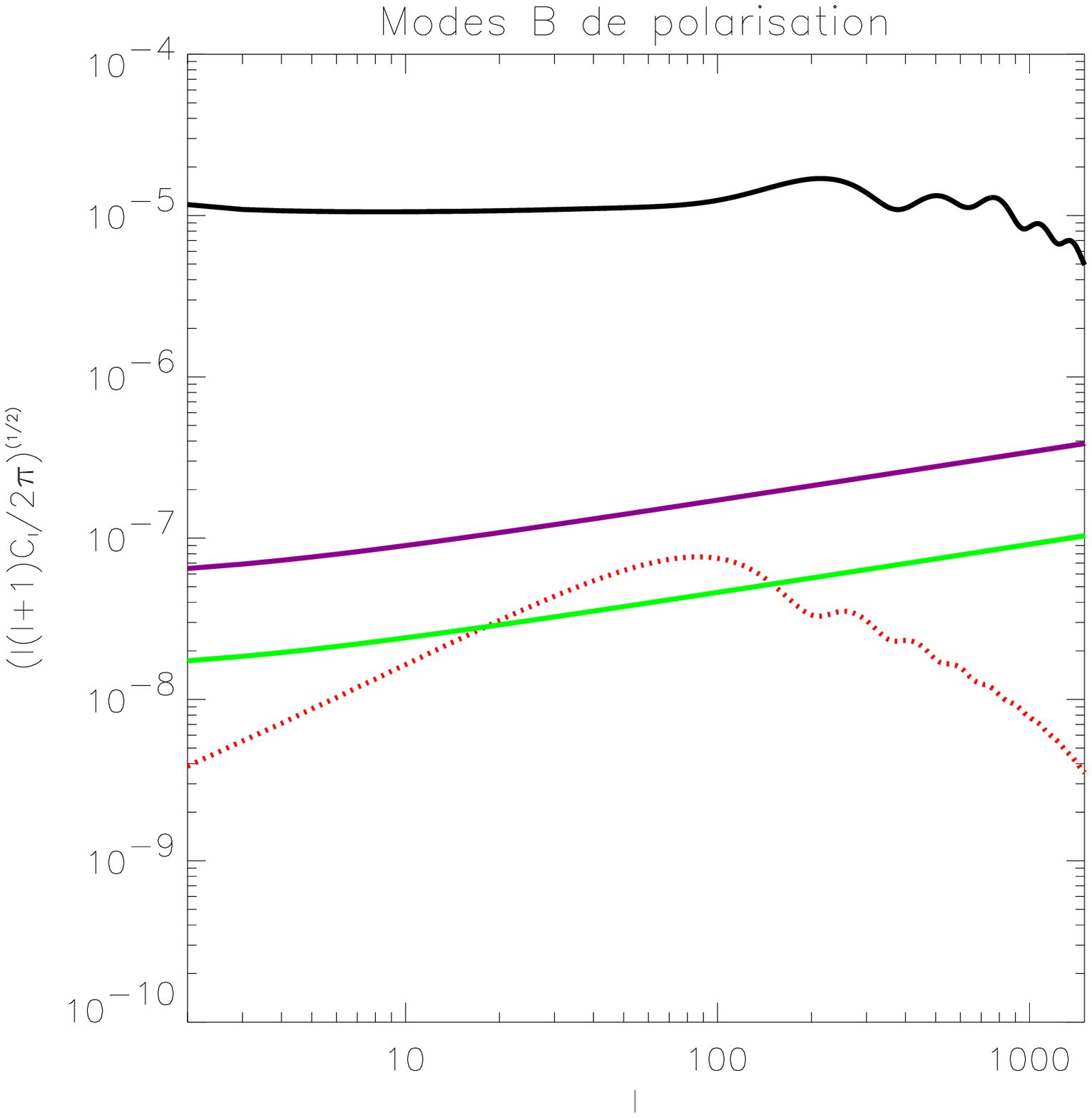}} 
\mbox{\includegraphics[width=\textwidth/3]{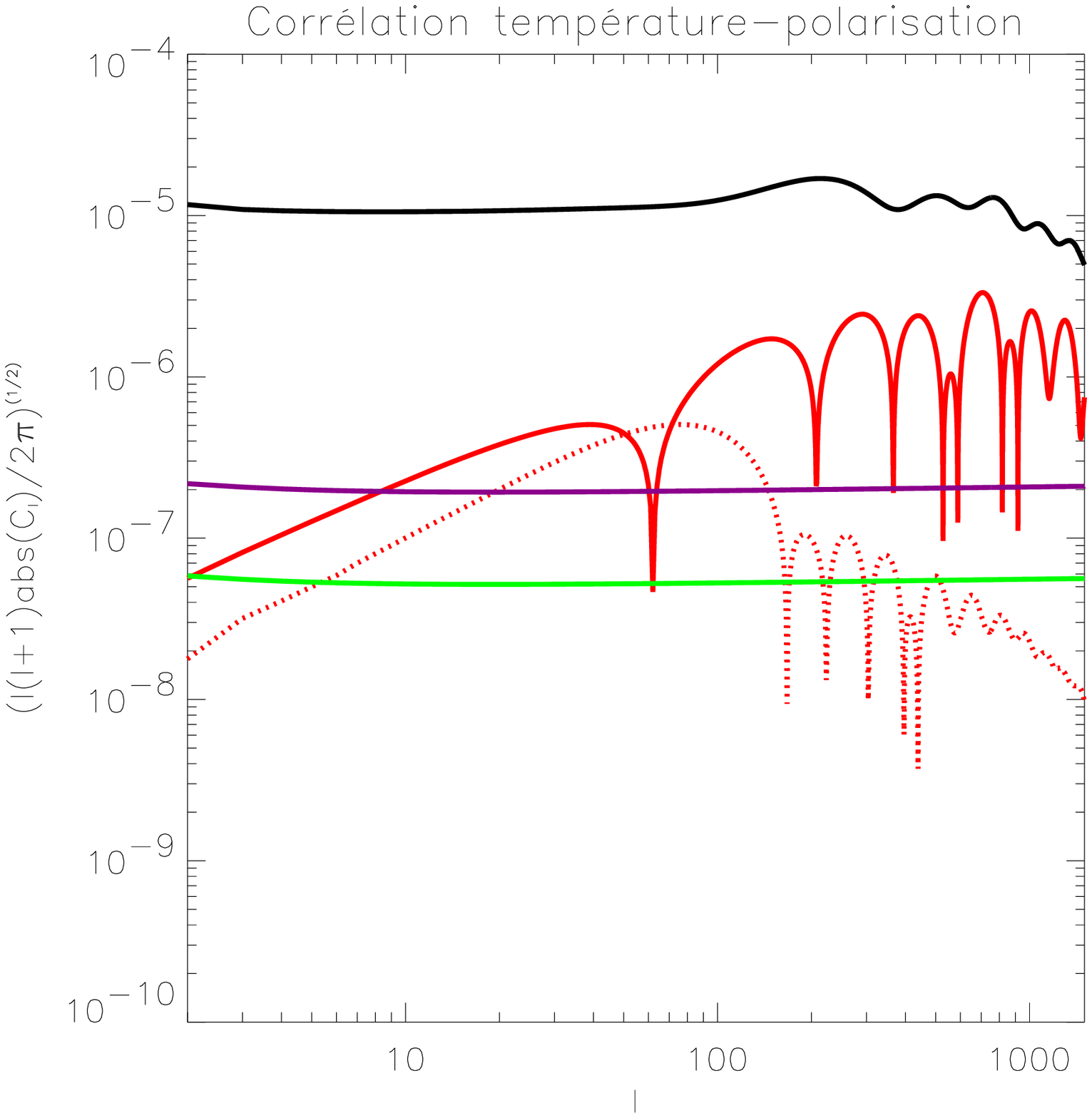}}
\end{tabular}
\caption{Predicted polarized $EE$, $TE$ and $BB$ power-spectra of Galactic dust,
for two Planck HFI channels at $143$ and $217\,{\rm GHz}$, compared to the predicted
spectra of the CMBP for a typical cosmological model. The green lines correspond to the $143\,{\rm GHz}$ 
dust spectra, and the purple ones to the same spectra at $217\,{\rm GHz}$. The red lines show the different 
CMBP spectra for the chosen cosmological model, and the black line gives the level of the temperature $TT$ power
spectrum for comparison.}
\end{figure}

As explained in the beginning of this section, the estimation of the contamination of CMBP measurements
by polarized dust emission requires not only the knowledge of the line of sight statistics of the polarization
degree of dust emission, but also the spatial distribution of this emission.
This in turn requires some knowledge of the 3-dimensional distribution of the dust and magnetic field lines.
Prunet et al. (1998) argued that the Galactic HI maps (with their velocity-space information), together with the 
observed strong correlation of HI gas and dust distribution for low column densities (Boulanger et al. 1996) could
be used to estimate power spectra of polarized dust emission. 
It should be noted here that their method relies on very simplistic assumptions to relate the velocity-space structure
of the HI gas to the 3d distribution of the gas, as well as simple extreme cases of cross-correlation of the magnetic field
distribution with the underlying gas density; so that the inferred power spectra represent at best a guideline to the expected
distribution of dust polarized emission.
However, the different assumptions concerning the interplay between the magnetic lines orientations and the gas density
structures are wide enough that this rough modelling should provide good upper and lower bounds on the 
slope/normalization of the polarized dust emission power spectra.
In order to estimate a ``worst case scenario'' of CMBP measurement contamination, they assumed a perfect alignment
of dust grains on magnetic field lines in the diffuse ISM, so that the problem becomes independent of the magnetic
field amplitude, as well as the dust temperature. This is of course not expected to be true for realistic
grain alignment mechanisms, be it paramagnetic relaxation of radiative torques.
With these assumptions they considered three cases for the relation between magnetic field lines and gas
density:
\begin{itemize}
\item{magnetic field lines parallel to gas filaments, defined as least gradient directions in the HI distribution.}
\item{magnetic field lines randomly distributed in the plane perpendicular to these directions. This could represent the 
case of helical field configurations.}
\item{constant magnetic field lines orientation.}
\end{itemize}
They applied their method to the HI Dwingaloo survey at intermediate Galactic latitude to compute an average
contamination for all-sky CMBP measurements. The first statistics that they derived is the histogram of the polarization 
degree in emission for the three different assumptions above, shown in figure~\ref{fig:histogram}.
They used the same method to predict the polarized (spatial) power spectra of dust emission, more specifically the power
spectra of the \emph{electric} (E) and \emph{magnetic} (B) modes of polarization, in order to compare them to the theoretical
predictions for the CMBP emission (see figure~\ref{fig:spectra}).
The power spectra for the electric and magnetic modes of polarized dust emission were computed in the flat sky 
approximation (Seljak 1996) from the Fourier coefficients of the Stokes parameters Q and U maps. 
They were then fitted by power laws, together with the temperature-polarization cross-correlation spectrum.
This decomposition in electric and magnetic modes of polarization is of particular importance for the CMBP emission
since scalar primordial perturbations of the metric can only produce electric modes of polarization in the CMBP,
thus defining the magnetic modes as a tracer of the primordial tensor (gravitational waves) perturbations (see Seljak \&
Zaldarriaga 1997, Kamionkowski et al. 1997).
The figures show the expected level of contamination by dust polarized emission for the two polarized channels of the 
Planck High Frequency Instrument that are most sensitive to CMBP ($143$ and $217 {\rm GHz}$). 
One can see that for a broad range of scales the measurement of electric modes of polarization in the CMBP should not be
too much hindered by Galactic dust emission. This is even more true for the $TE$ cross-correlation.
The B modes however (shown here for a typical cosmological model with $n_T=-0.1$ and $T/S=-7n_T$) are seriously
contaminated by the dust emission, and our ability to measure them will rely heavily on our capacity of using the multi-frequency
measurements of the polarized diffuse emissions in the millimeter and sub-millimeter wavelengths to disentangle the cosmological
and Galactic signals (see e.g. Bouchet et al. 1999, Tegmark et al. 2000) using the fact that the intrinsic polarization degree
in emission in the submillimeter should be roughly independent of wavelength.

This task could however be complicated by the expected complexity of the dust polarized emission (in particular the possible
dependence of polarization efficiency with dust temperature, and more generally on the grains environment.
It is therefore of prime importance for cosmology to better understand the polarized emission of Galactic dust and its behaviour,
both spatially and spectrally.

\section{Summary}

The principal points discussed above are as follows:

\begin{itemize}

\item Dust provides the most intricate pattern of polarized radiation.
The dependence of polarization of grain temperature, composition,
size and environment makes the use of templates difficult.

\item If anomalous emission in the range of  10-100~GHz is due to spinning
dust particles, the polarization of the emission is marginal for
frequencies larger than $\sim 35$~GHz. If the anomalous emission or
part of it is due to magneto-dipole mechanism the polarization may
be substantial and may exhibit reversals of direction with frequency.

\item To get a better insight into the microwave properties of dust
more laboratory studies are necessary. Some of them, e.g. measurements
of the magnetic susceptibility of candidate materials at microwave 
frequencies, are trivial using the modern technology.

\item Systematic studies of microwave polarization arising 
from dust will enable to determine the pattern of the
CMB polarization and will shed light on
many problems, including those of interstellar magnetic field and dust
composition.
 
\end{itemize}

\begin{theacknowledgments}
AL is thankful to John Mathis for useful exchanges.
AL acknowledges the support of NSF Graint AST-0125544. 
\end{theacknowledgments}

\end{document}